\documentclass[aps,pra,showpacs,twocolumn,superscriptaddress]{revtex4}

\usepackage{graphicx,color}
\usepackage{amsmath,amsthm,amsfonts,amssymb,bm}
\usepackage{times}
\usepackage{epsf}
\usepackage[colorlinks={true}]{hyperref}

\usepackage[T1]{fontenc}
\usepackage[utf8]{inputenc}
\usepackage{tabularx,ragged2e,booktabs,caption}
\newcolumntype{C}[1]{>{\Centering}m{#1}}

\hypersetup{citecolor={blue}, filecolor={blue}, linkcolor={blue}, urlcolor={blue}}

\newcommand{\commentold}[1]{}

\begin{document}

\title{Probing the degree of non-Markovianity for independent and common environments}

\author{Felipe F. Fanchini}
\affiliation{Faculdade de Ci\^encias, UNESP - Universidade Estadual Paulista, Bauru, SP, 17033-360, Brazil}
\author{G\"{o}ktu\u{g} Karpat}
\affiliation{Faculty of Engineering and Natural Sciences, Sabanci University, Tuzla, Istanbul, 34956, Turkey}
\author{Leonardo K. Castelano}
\affiliation{Departamento de F\'{\i}sica, Universidade Federal de S\~ao Carlos, S\~ao Carlos, SP, 13565-905, Brazil}
\author{Daniel Z. Rossatto}
\affiliation{Departamento de F\'{\i}sica, Universidade Federal de S\~ao Carlos, S\~ao Carlos, SP, 13565-905, Brazil}

\date{\today}

\begin{abstract}
We study the non-Markovianity of the dynamics of open quantum systems focusing on the cases of independent and common environmental interactions.
We investigate the degree of non-Markovianity quantified by two distinct measures proposed by Luo, Fu and Song (LFS) and Breuer, Laine and Pillo (BLP).
We show that the amount of non-Markovianity, 
for a single and a pair of qubits, depends on the quantum process, the proposed measure and whether the environmental interaction is collective or independent. In particular, we demonstrate that while the degree of non-Markovianity generally increases with the number of the qubits in the system for independent environments, the same behavior is not always observed for common environments. In the latter case, our analysis suggests that the amount of non-Markovianity could increase or decrease depending on the properties of the considered quantum process.
\end{abstract}


\maketitle

\section{Introduction}
The concept of non-Markovianity is a prominent aspect of the dynamics of open quantum systems and has been attracting both theoretical and experimental attention in the last few years~\cite{general,haikkabec}. Moreover, it has been shown that non-Markovianity can be used as a tool in quantum protocols \cite{bogna}, can be employed to take advantage in quantum metrology~\cite{metro}, and can be exploited in quantum key distribution~\cite{key}.
Further concepts behind the non-Markovian dynamics have also been investigated, for example, the influence of environment size \cite{size} and the possibility to pursue new quantum technologies by using non-Markovian effects~\cite{bogna}. Although all those efforts to understand the connection between non-Markovian dynamics and quantum information theory have been performed, measuring non-Markovianity is complicated and generally only small systems have been considered~\cite{onetwo}.
{However, the real usefulness of a quantum system for computation or simulation is only appreciable in the limit of large-scale information processing. Therefore, it is fundamental to understand the properties of non-Markovianity for multipartite systems.} 

Recently, various measures for quantifying the degree of non-Markovianity of the dynamics of an open quantum system \cite{BLP, RHP, Luo, other} have been introduced in the literature; however, there is no consensus on what precisely determines the non-Markovianity of a dynamical quantum process. It has been demonstrated that the conclusions drawn from different measures might not agree depending on the considered physical model. The most widely used measure of non-Markovianity has been introduced by Breuer, Laine and Piilo (BLP) \cite{BLP}. In their seminal paper, they claimed that information flows only from the system into the environment for a Markovian process and the information flow can be measured by the trace distance of two arbitrary quantum states, which probes the distinguishably between them. To implement this measure, one needs to perform an optimization by checking the dynamics of the trace distance for a huge number of initial sates. Thus, this procedure is very demanding and almost impracticable when dealing with multipartite systems. Moreover, Rivas, Huelga and Plenio (RHP) have constructed a measure of non-Markovianity that quantifies the deviation from divisibility for a dynamical map \cite{RHP}, which is also difficult to be implemented in general. 
In order to overcome these difficulties, we use an efficient method to evaluate a measure of non-Markovianity, recently proposed by Luo, Fu, and Song (LFS), based on the non-monotonical behavior of the quantum mutual information for non-Markovian processes \cite{Luo}. This measure coincides with other important ones such as the BLP measure for quite general cases and can be straightforwardly extended for studying multipartite systems.

When dealing with the interaction between a quantum system and its environment, there are two important physical processes that must be considered: relaxation and decoherence (here called dephasing). While relaxation is associated to a process involving loss of energy, dephasing is associated to the loss of purity without any exchange of energy between the system and its surroundings. 
In this work, we explore both processes considering the amplitude damping channel to describe dissipative processes, and taking into account two different kinds of interactions to describe phase damping processes, namely a superohmic dephasing channel and the  phase damping case employed to describe impurity atoms coupled to a Bose-Einstein condensate. For all different scenarios, we analyze the effect of both independent and common environmental interactions on the behavior of non-Markovianity 
by investigating two distinct quantifiers of the degree of non-Markovianity given by the LFS and the BLP measures. For zero temperature environments, we link the LFS measure to the rate of change of the system entropy $S(\rho^s(t))$ and the environment entropy $S(\rho^e(t))$. We show that, for the LFS measure, a quantum process is non-Markovian if the time derivative of $S(\rho^s(t))$ is greater than the time derivative of $S(\rho^e(t))$. We present a detailed analysis of the evaluation of the LFS measure for a single qubit, and discuss the behavior of the optimal initial states as a function of the parameters of the environments. Moreover, we demonstrate that the degree of non-Markovianity, for both proposed measures, \textit{increases} in general as a function of the number of qubits in the system for the case of independent environments. On the other hand, for global environments, we show that the amount of non-Markovianity depends on the quantum process and the proposed measure.

The paper is organized as follows. In Section II, we introduce the measures of non-Markovianity that will be used in our investigation. Section III outlines several system-environment models describing the dynamics of open quantum systems. Section IV and Section V include our findings related to the behavior of the LFS and BLP measures under the considered models, respectively. Section VI covers the summary of the results obtained in this work.

\section{Measuring non-Markovianity}
\subsection{LFS MEASURE}
The definition of the LFS measure of non-Markovianity \cite{Luo} is based on the following: suppose that we have a quantum system in a Hilbert space $H$, and a quantum process $\Lambda(t)$ governing the dynamical evolution of the considered system. If an arbitrary ancilla system in a Hilbert space $H^a$ is introduced, the composite state of the main system and ancilla $\rho^{sa}$ pertains to the Hilbert space $H \otimes H^a$. In this case, assuming a trivial dynamics on the ancillary, the time evolution of the total system is given by $\rho^{sa}(t)=(\Lambda(t)\otimes I) \rho^{sa}(0)$, where $I$ is the identity operator acting on the state space of the ancillary. The amount of total correlations in a bipartite state $\rho^{sa}$ can be quantified through the quantum mutual information
\begin{equation}
I(\rho^{sa})=S(\rho^s)+S(\rho^a)-S(\rho^{sa}),\label{mutual}
\end{equation}
where $\rho^a=\textmd{tr}_s \rho^{sa}$ and $\rho^s=\textmd{tr}_a \rho^{sa}$ represent the reduced density operators of the system and the ancilla, respectively. $S(\rho)=-\textmd{tr}\rho \log_2\rho$ is the von Neumann entropy. 
Exploiting the fact that quantum mutual information decreases monotonically as a function of time for a Markovian process, LFS have proposed a new quantity for measuring the non-Markovianity of the dynamical process $\Lambda(t)$ from an informational perspective:
\begin{equation}
\mathcal{N}(\Lambda)=\sup_{\rho^{sa}(0)}\int_{(d/dt)I(\rho^{sa}(t))>0}\frac{d}{dt}I(\rho^{sa}(t))dt,\label{N}
\end{equation}
where the $\sup$ is taken over all possible initial states $\rho^{sa}(0)$. Even though this measure has an interesting meaning for the quantification of non-Markovianity, its evaluation is hard to perform due to the potentially complex optimization problem.

As described in Eq. (\ref{N}), a dynamical quantum process is said to be non-Markovian if $\frac{d}{dt}I(\rho^{sa}(t))>0$. Note that the ancilla, unlike the system, does not interact with the environment. In other words, the state of the ancilla is time independent. Therefore, the time derivative of the quantum mutual information can be written as
\begin{equation}
\frac{d}{dt}I(\rho^{sa}(t)) = \frac{d}{dt}S(\rho^s(t)) - \frac{d}{dt}S(\rho^{sa}(t)).
\end{equation}
For a zero temperature environment, an interesting result can be obtained from this equation. Since we take $\rho^{sa}(0)$ as a pure state, and the environment starts in the state $\rho^e(0)=|0\rangle\langle0|$ for the zero temperature case, the total quantum state composed of the system, the ancilla and the environment is a pure state at any time, leading to $S(\rho^{sa}(t))=S(\rho^e(t))$. Consequently, following the LFS measure, we obtain a non-Markovianity criterion without the need of an ancilla; therefore, a quantum process is non-Markovian if and only if
\begin{equation}
\frac{d}{dt}S(\rho^s(t)) > \frac{d}{dt}S(\rho^e(t)).
\end{equation}
This condition links the non-Markovianity measure to the rate of change of the system and the environment entropies. Nonetheless, some peculiar aspects should be noted: the environment is initially in a pure state by assumption, but no restrictions were imposed to the system. If the system is initially in a pure state as well, we have $\rho^{se}$ pure and therefore $S(\rho^s(t))=S(\rho^e(t))$. Such a result does not mean that the process is actually Markovian because we need to maximize over all possible initial conditions of the system to be able to determine the degree of non-Markovianity. An equivalent equation for Eq. (\ref{N}) can be deduced without the necessity of an ancilla:
\begin{equation}
\mathcal{N}_{T=0K}(\Lambda)=\sup_{\rho^{s}(0)}\int_{(d/dt)\Delta S_{se}(t)>0}\frac{d}{dt}\Delta S_{se}(t)dt,\label{NT0}
\end{equation}
with $\Delta S_{se}(t) = S(\rho^{s}(t))-S(\rho^{e}(t))$. The advantage of the above equation over Eq. (\ref{N}) is the fact that the maximization is just over the initial conditions of the system instead of the initial conditions of the composite state of the system and the ancilla, which is required to calculate $\mathcal{N}(\Lambda)$.
It is important to emphasize that the time derivative of the entropy is the important quantity for a non-Markovian process. Furthermore, to calculate $S(\rho^e(t))$, we do not need to worry about the state of the environment. The idea here is to maximize $\mathcal{N}_{T=0K}$ over all possible initial system states and, for each choice, purify it including an extra subsystem. Because the environment is set in a pure state at $t=0$, the entropy of the system plus the purifier subsystem is equal to the entropy of the environment at any time. An important observation that deserves to be mentioned is that, to evaluate Eq. (\ref{N}) and Eq. (\ref{NT0}), it is possible to suppress the calculation of the integrals and time derivatives of the integrands. In fact, it is straightforward to note that we can rewrite the LFS measure as:
\begin{equation}
\mathcal{N}(\Lambda)=\sup_{\rho^{sa}(0)}\sum_i\left[ I\textbf{(}\rho^{sa}(b_i)\textbf{)} - I\textbf{(}\rho^{sa}(a_i)\textbf{)} \right].\label{Nsum}
\end{equation}
To compute this quantity, we first determine the time intervals $(a_i,b_i)$ in which the mutual information increases, then we sum up the contribution of each interval to obtain $\mathcal{N}(\Lambda)$. 

On the other hand, LFS presented a significant simplification for Eq. (\ref{N}) in Ref. \cite{Luo}. Assuming that $H^a = H$ and $\rho^{sa}(0)=|\Psi\rangle \langle \Psi|$ where $|\Psi\rangle$ is \textit{any} maximally entangled pure state of the system and the ancilla, they obtain an easily computable measure of non-Markovianity:
\begin{equation}
\mathcal{N}_0(\Lambda)=\int_{(d/dt)I(\rho^{sa}(t))>0}\frac{d}{dt}I(\rho^{sa}(t))dt,\label{luomeas}
\end{equation}
with $\rho^{sa}(t)=(\Lambda(t)\otimes I)|\Psi\rangle \langle \Psi|$. In Appendix A, we explicitly show that $\mathcal{N}_0(\Lambda)$, despite its utility as a witness for non-Markovianity, may be misleading and give an inaccurate conclusion about the degree of non-Markovianity of a quantum process. Furthermore, we demonstrate that $\mathcal{N}(\Lambda)$ does not depend on the amount of entanglement shared between the system and ancilla because two distinct initial states with the same degree of entanglement can give different results. Actually, the optimal state is not maximally entangled in general.
\subsection{BLP MEASURE}

The BLP measure of non-Markovianity \cite{BLP} employs the trace distance $D_{12}(t)=1/2 \rm{tr}|\rho_1(t)-\rho_2(t)|$ between two arbitrary reduced density matrices $\rho_1(t)$ and $\rho_2(t)$ in order to check the distinguishably between them. Such reduced density matrices are resulting from the calculation of the partial trace with respect to the environment part of the total density matrix that describes a system coupled to an environment. When $dD_{12}(t)/dt<0$, the distinguishability between the reduced density matrices decreases and there is a flow of information from the system to the environment. On the other hand, the information flow from the environment back to the system can happen if $dD_{12}(t)/dt>0$. In such situations, the processes are said to be non-Markovian. Furthermore, the BLP measure of non-Markovianity \cite{BLP} is mathematically defined as follows:
\begin{eqnarray}
\mathcal{N}_{BLP}(\Lambda)&=&\max_{\rho_1(0),\rho_2(0)}\int_{(dD_{12}(t)/dt)>0}\frac{dD_{12}(t)}{dt}dt \label{blpmeas} 
\end{eqnarray}
where 
the maximum value is taken over all pairs of initial reduced states $\rho_1(0)$ and $\rho_2(0)$.
It is important to emphasize that to numerically implement this measure, it is necessary to evaluate the dynamics of the trace distance $D_{12}(t)$ for a huge number of initial sates which makes this procedure very demanding when dealing with multipartite systems.
We note that the above equation can also be rewritten as
\begin{equation}
\mathcal{N}_{BLP}(\Lambda)=\max_{\rho_1(0),\rho_2(0)}\sum_i [D_{12}(b_i)-D_{12}(a_i)],\label{Sblpm}
\end{equation}
where time intervals $(a_i,b_i)$ correspond to the regions in which $dD_{12}(t)/dt>0$ and
the maximum value is taken over all pairs of initial reduced states $\rho_1(0)$ and $\rho_2(0)$.


\section{Open System Dynamics}

\subsection{Phase Damping Channel}

We consider a spin-boson type Hamiltonian $H_{IPD}$ that describes a pure dephasing type of interaction between a qubit and a bosonic environment:
\begin{equation}
H_{IPD}=\frac{\omega_0}{2}\sigma_z+\sum_{k}\omega_k a_k^\dagger a_k+\sum_{k}\sigma_z(g_k a_k^\dagger + g_k^* a_k),\label{HIPD}
\end{equation}
where the first and the second terms of Eq. (\ref{HIPD}) are responsible for the free evolution of the qubit and the environment, respectively. The third term of Eq. (\ref{HIPD}) accounts for the interaction between the qubit and its environment. We first note that $[H,\sigma_z]=0$, which immediately implies the absence of transitions between different energy levels. Thus, the population terms in the density matrix of the system are conserved quantities. Here, $\omega_0$ is the transition frequency of the qubit and $\omega_k$ is the field frequency of the $k$-th environmental field mode. The constant $g_k$ controls the strength of the coupling between the qubit and each field mode of the environment. While the qubit operator is given by the usual Pauli $\sigma_z$ matrix, the creation operator $a_k$ and the annihilation operator $a_k^\dagger$, satisfying the bosonic commutation relations $[a_k,a_{k'}^\dagger]=\delta_{k,k'}$, represent the environment. It is worth to stress that this qubit plus environment model admits an exact solution \cite{soldeph}. 
We assume that the composite state of the qubit and the environment are initially factorized, that is, there exist no correlations between the system and the environment at $t=0$; furthermore, the environment is initially in its vacuum state $\rho^e(0)=|0\rangle\langle0|$ at zero temperature. We consider a sufficiently large environment; therefore, we can replace the sum over the discrete coupling constants by an integral over a continuous distribution of frequencies of the environmental modes, \textit{i.e.}, $\sum_k |g_k|^2 \rightarrow \int_0^\infty d\omega J(\omega)$. In addition, we suppose that the spectral density of the environmental modes is Ohmic-like
\begin{equation}
J(\omega)=\eta\frac{\omega^s}{\omega_c^{s-1}}e^{-\omega/\omega_c}\label{spect},
\end{equation}
with $\omega_c$ being the cut-off frequency and $\eta$ a dimensionless coupling constant. Depending on the parameter $s$, the spectral density is called subohmic $(s<1)$, ohmic $(s=1)$ or superohmic $(s>1)$. Under these conditions, the dynamics of a single qubit can be obtained in the operator-sum representation as
\begin{equation}
\rho(t)=\sum_{i=1}^{2} K_i(t) \rho(0) K_i^\dagger(t),
\end{equation}
where the Kraus operators $K_i(t)$ are given by
\begin{align}
  K_1(t) &= \begin{pmatrix} 1 & 0\\ 0 & r(t) \end{pmatrix}, &
  K_2(t) &= \begin{pmatrix} 0 & 0\\ 0 & \sqrt{1-r^2(t)} \end{pmatrix},
\end{align}
with $\sum_{i=1}^{2} K_i^\dagger(t) K_i(t) = I$ for all values of $t$, where $I$ denotes the $2\times2$ identity matrix. Here, the dephasing parameter $r(t)$ is
\begin{equation}
r(t)=\exp\left[-\int_0^t \gamma(t')dt'\right],
\end{equation}
where the dephasing rate $\gamma(t)$ takes the form
\begin{equation}
\gamma(t)=\eta \omega_c (1+(\omega_c t)^2)^{-s/2}\Gamma(s)\sin(s\arctan(\omega_c t)),
\end{equation}
with $\Gamma(s)$ being the Euler gamma function.


\subsection{Amplitude Damping Channel}
In order to discuss the relaxation process, we consider the following model Hamiltonian
\begin{equation}
H_{IAD}=\omega_0\sigma_{+}\sigma_{-}+\sum_{k}\omega_k a_k^\dagger a_k + (\sigma_{+}B + \sigma_{-}B^\dagger)\label{had},
\end{equation}
where $B=\sum_k g_k a_k$ with $g_k$ being the coupling constant. The first two terms of Eq. (\ref{had}) describe the free evolution of the qubit and the environment, respectively, while the third term accounts for the interaction between the qubit and the environment. The transition frequency of the qubit is $\omega_0$, and $\sigma_{\pm}$ denotes the raising and lowering operators related to the qubit. The index $k$ is used to label the different environmental field modes with frequencies $\omega_k$, which are mathematically described by the annihilation and creation operators given by $a_k$ and $a_k^\dagger$, respectively. 
Restricting ourselves to the case of a single excitation, the modes of the environment can be described by an effective spectral density of the form
\begin{equation}
J(\omega)= \frac{1}{2\pi}\frac{\gamma_0 \lambda^2}{(\omega_0 - \omega)^2 + \lambda^2},
\end{equation}
where $\lambda$ defines the spectral width of the coupling and it is also connected to the correlation time of the environment $\tau_B$ by the relation $\tau_B\approx 1/\lambda$. $\gamma_0$ is the time scale $\tau_R$ over which the state of the system changes by $\tau_R\approx 1/\gamma_0$. For this form of a spectral density, it is not hard to distinguish the weak and the strong coupling regimes. The case $\tau_R > 2\tau_B$ corresponds to the weak coupling regime where the decoherence process is Markovian because the relaxation time is greater than the correlation time of the environment. On the other hand, the case $\tau_R <2\tau_B$ corresponds to the strong coupling regime where the non-Markovian nature of the environment becomes evident. We note that at zero temperature this Hamiltonian with the considered spectral density (known as the damped Jaynes-Cummings model in the literature) represents one of the few exactly solvable models for open quantum systems. In the strong coupling regime, the time evolution of a single qubit can be expressed in the operator-sum representation as
\begin{equation}
\rho(t)=\sum_{i=1}^{2} M_i(t) \rho(0) M_i^\dagger(t),
\end{equation}
where the corresponding Kraus operators $M_i(t)$ are given by
\begin{align}
  M_1(t) &= \begin{pmatrix} 1 & 0\\ 0 & \sqrt{p(t)} \end{pmatrix}, &
  M_2(t) &= \begin{pmatrix} 0 & \sqrt{1-p(t)}\\ 0 & 0 \end{pmatrix},
\end{align}
satisfying the condition $\sum_{i=1}^{2} M_i^\dagger(t) M_i(t) = I$ for all values of $t$. The damping parameter $p(t)$ reads
\begin{equation}
p(t)=e^{-\lambda t} \left[ \cos{\left(\frac{dt}{2}\right)+\frac{\lambda}{d}\sin{\left(\frac{dt}{2}\right)}}\right]^2,
\end{equation}
with $d=\sqrt{2\gamma_0\lambda-\lambda^2}$.
\vspace{0.5cm}

\subsection{Impurity Atoms Coupled to a Bose-Einstein Condensate}

The third model considered in this work deals with two atoms interacting with an ultracold bosonic
Rubidium gas in a Bose-Einstein condensate (BEC) state \cite{Cirone,sabrina2}.
Here, the qubit is represented by an impurity atom in a double-well potential of an
optical superlattice of wavelength $\lambda$, where the size of the qubit
is the distance between the lattice sites, $L=\lambda /4$. The superlattice
is immersed in the BEC environment, where the Rubidium gas
is assumed to be in the weak coupling regime, justifying the validity of Bogoliubov approach. For more details
on the model, see Ref. \cite{Cirone,sabrina2}.

In Ref. \cite{sabrina2}, the authors studied the non-Markovianity of this model
under the point of view of the BLP and RHP measures, and they showed
that it is possible to tune from a common environment to an
independent one by adjusting the spatial separation of the qubits. Moreover, the
authors observed that whereas the BLP measure is super-additive when the qubits are
very close to each other (common environment regime), it is sub-additive
when the qubits are sufficiently far enough from each other (independent environment
regime).

The dynamics of the model is given by a Lindblad-type master equation with
time-dependent decay rates \cite{Cirone,sabrina2}%
\begin{widetext}
\begin{eqnarray}
\frac{d\rho }{dt} &=&\frac{\gamma _{1}\left( t\right) -\gamma _{2}\left(
t\right) }{2}\left[ \left( \sigma _{z}^{(1)}-\sigma _{z}^{(2)}\right) \rho
\left( \sigma _{z}^{(1)}-\sigma _{z}^{(2)}\right) -\frac{1}{2}\left\{ \left(
\sigma _{z}^{(1)}-\sigma _{z}^{(2)}\right) \left( \sigma _{z}^{(1)}-\sigma
_{z}^{(2)}\right) ,\rho \right\} \right]\nonumber  \\
&&+\frac{\gamma _{1}\left( t\right) +\gamma _{2}\left( t\right) }{2}\left[
\left( \sigma _{z}^{(1)}+\sigma _{z}^{(2)}\right) \rho \left( \sigma
_{z}^{(1)}+\sigma _{z}^{(2)}\right) -\frac{1}{2}\left\{ \left( \sigma
_{z}^{(1)}+\sigma _{z}^{(2)}\right) \left( \sigma _{z}^{(1)}+\sigma
_{z}^{(2)}\right) ,\rho \right\} \right] ,\label{becdyna}
\end{eqnarray}\end{widetext}
where $\sigma _{z}^{(n)}$ is the usual Pauli matrix for the $n$-th atom $\left(
n=1,2\right)$, and%
\begin{widetext}
\begin{eqnarray}
\gamma _{1}\left( t\right)  &=&\frac{g_{SE}^{2}n_{0}}{\hbar \pi ^{2}}%
\int_{0}^{\infty }dkk^{2}e^{-k^{2}\sigma ^{2}/2}\frac{\sin \left( \frac{E_{k}%
}{2\hbar }t\right) \cos \left( \frac{E_{k}}{2\hbar }t\right) }{\left(
\epsilon _{k}+2g_{E}n_{0}\right) }\left( 1-\frac{\sin \left( 2kL\right) }{2kL%
}\right)  \\
\gamma _{2}\left( t\right)  &=&\frac{g_{SE}^{2}n_{0}}{2\hbar \pi ^{2}}%
\int_{0}^{\infty }dkk^{2}e^{-k^{2}\sigma ^{2}/2}\frac{\sin \left( \frac{E_{k}%
}{2\hbar }t\right) \cos \left( \frac{E_{k}}{2\hbar }t\right) }{\left(
\epsilon _{k}+2g_{E}n_{0}\right) }\left( \frac{\sin \left( 2k\left(
D+L\right) \right) }{2k\left( D+L\right) }+\frac{\sin \left( 2k\left(
D-L\right) \right) }{2k\left( D-L\right) }-2\frac{\sin \left( 2kD\right) }{%
2kD}\right) ,\label{gam2}
\end{eqnarray}%
\end{widetext}
where $g_{E}=4\pi \hbar ^{2}a_{E}/m_{E}$ is the boson-boson coupling for a
BEC environment with scattering length $a_{E}$ and atomic mass $m_{E}$, $
g_{SE}=2\pi \hbar ^{2}a_{SE}/m_{SE}$ is the coupling between the system and
the environment with scattering length $a_{SE}$ and reduced mass $
m_{SE}=m_{S}m_{E}/\left( m_{S}+m_{E}\right) $. $E_{k}=\sqrt{2\epsilon
_{k}n_{0}g_{E}+\epsilon _{k}^{2}}$ is the energy of the $k$-th Bogoliubov
mode, $n_{0}$ is the condensate density, $\epsilon _{k}=\hbar
^{2}k^{2}/2m_{E}$ and $\sigma $ is the variance parameter of the lattice
site. Finally, $2D\geq 8L$ is the distance between the atoms. Similarly to Ref.
\cite{sabrina2}, we consider $^{23}\rm{Na}$ impurity atoms immersed in a $^{87}\rm{Rb}$
condensate with $\lambda =600$ nm and $n_{0}=10^{20}$ m$^{-3}$.
The scattering length of the atoms is $a_{Rb}=99a_{0}$ where $a_{0}$ is
the Bohr radius, and we assume $a_{SE}=55a_{0}$. Finally, we choose the scattering
length of the BEC environment as $a_E=0.5a_{Rb}$.

\section{Degree of Non-Markovianity: LFS Measure}

\subsection{Single qubit}

Before starting to elucidate the properties of the LFS measure for multipartite systems at zero temperature, we consider the case of a single qubit system. In this case, the maximization in Eq. (\ref{NT0}) can be numerically evaluated because the general form of the density matrix $\rho^s(0)$ depends just on three real variables. Explicitly,
\begin{equation*}
\rho^s(0)=
\begin{pmatrix}
\rho_{11}(0) & \Re[\rho_{12}(0)]+i\Im[\rho_{12}(0)]\\
\Re[\rho_{12}(0)]-i\Im[\rho_{12}(0)] & 1-\rho_{11}(0)
\end{pmatrix}.
\end{equation*}
For both dephasing processes, our numerical analysis show that the maximum in Eq. (\ref{N}) is reached for a maximally mixed initial state. In other words, the optimal state of the composite system $\rho^{sa}(0)$ is maximally entangled, justifying the simplification proposed by LFS, as described in Eq. (\ref{luomeas}). However, contrarily to what one might expect, the maximum value in Eq. (\ref{N}) for the relaxation process is obtained for a diagonal initial state whose system plus ancilla density matrix is \textit{not} maximally entangled.

As suggested by our numerical investigation, we first set zero the off-diagonal elements of the density matrix, \textit{i.e.} $\rho_{12}(0)=\rho_{21}(0)=0$. In Fig. (1-a) and Fig. (1-c) we plot the possible values of the degree of non-Markovianity quantified by the LFS measure ${\mathcal N(\Lambda)}$ for the superohmic PD channel with $s=3$, $w_c=1$ and $\eta=2$, and for the BEC environment with $\sigma=45$ nm, respectively, as a function of the density matrix population $\rho_{11}(0)$. As can be seen from the Fig. (1-a) and Fig. (1-c), the LFS measure corresponds to having $\rho_{11}(0)=0.5$, implying that the optimal initial composite state is maximally entangled. On the other hand, Fig. (1-b) presents the results of the same analysis performed for the AD channel with $\gamma_0=1$ and $\lambda=0.1$. It is shown that the LFS measure is obtained for $\rho_{11}(0)\approx0.4$, meaning the optimal composite state is not maximally entangled in this case. In particular, such a result points out that the LFS measure does not generally depend on the initial entanglement between the system and the ancilla. 

Another interesting point is related to the dependence of the optimal initial state of the system on the parameters of the considered environmental model. In Fig. (2-a), Fig. (2-b) and Fig. (2-c) we plot the density matrix element $\rho_{11}(0)$ of the optimal initial state for the superohmic PD channel, AD channel, and the BEC environment, respectively, as a function of the bath parameters $\omega_c$, $\lambda$, and $\sigma$. The results of this analysis demonstrate an important point, that is, whereas the optimal initial state for the superohmic PD channel and the BEC environment do not depend on the bath parameters $\omega_c$ and $\sigma$, the AD channel is highly sensitive to the bath parameter $\lambda$. In fact, we have found that as the parameter $\lambda$ gets smaller, the optimal state tends to be a maximally mixed one.

\begin{figure}[htbp]
\includegraphics[width=.42\textwidth]{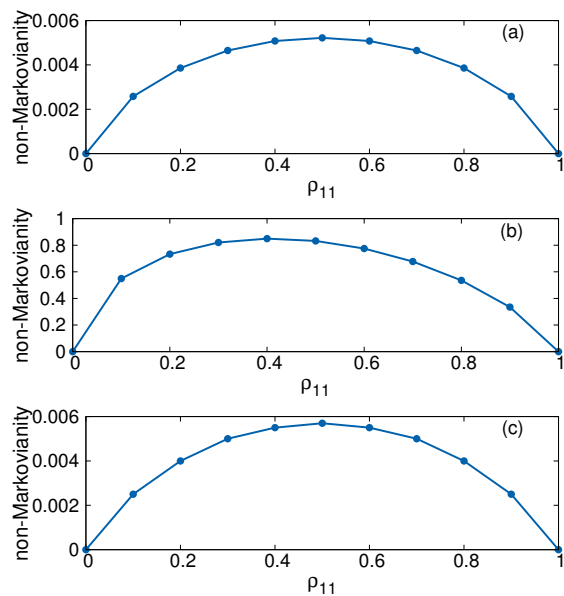} {}
\caption{Non-Markovianity for one qubit as a function of the density matrix element $\rho_{11}(0)$ for (a) the superohmic dephasing process with $s=3$, $w_c=1$ and $\eta=2$, for (b) the relaxation process with $\gamma_0=1$ and $\lambda=0.1$, and for (c) an impurity atom coupled to a BEC environment with $\sigma=45$ nm.}
\end{figure}

\begin{figure}[htbp]
\includegraphics[width=.42\textwidth]{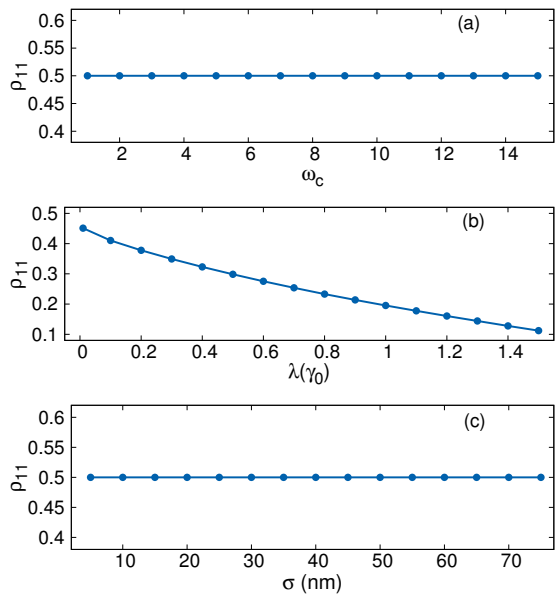} {}
\caption{The density matrix element $\rho_{11}(0)$ of the optimal initial state of a single qubit system as a function of the bath parameters for (a) the superohmic dephasing process with $s=3$ and $\eta=2$, for (b) the relaxation process, and for (c) an impurity atom coupled to a BEC environment.}
\end{figure}

\subsection{Multi-Qubit: Independent Environments}

In this section, we apply the LFS measure to the multipartite case taking into account only independent environments. First of all, we consider a quantum state which is composed of a system of $n$ qubits and an ancilla that purifies the system state. Assuming that only $n$ qubits are subjected to the environmental noise, and the ancillary system evolves freely, the dynamics of the composite system $\rho^{sa}(t)$ can be obtained, for the AD and superohmic PD channels, as
\begin{equation}
\rho^{sa}(t)=\sum_{i= \{1,2\}} (E_i^{\otimes n} \otimes I) \rho^{sa}(0) (E_i^{\otimes n} \otimes I)^\dagger,
\end{equation}
where $E_i$ are the Kraus operators describing the AD or superohmic PD channels for a single qubit, $n$ is the number of qubits, $I$ denotes the identity matrix with dimensions of the ancillary system, and the sum over the index ${i= \{1,2\}}$ runs over all possible permutations of the Kraus operators $E_i^{\otimes n}$. For the case of $n$ atoms independently coupled to a BEC environment, on the other hand, we evaluate the dynamics numerically extending Eq. (\ref{becdyna}) under the assumption that $D/L\rightarrow\infty$ in Eq. (\ref{gam2}). In this situation, since the dynamics of each atom are independent of each other, it is straightforward to consider more than two atoms.

\begin{figure} [htbp]
\includegraphics[width=.42\textwidth]{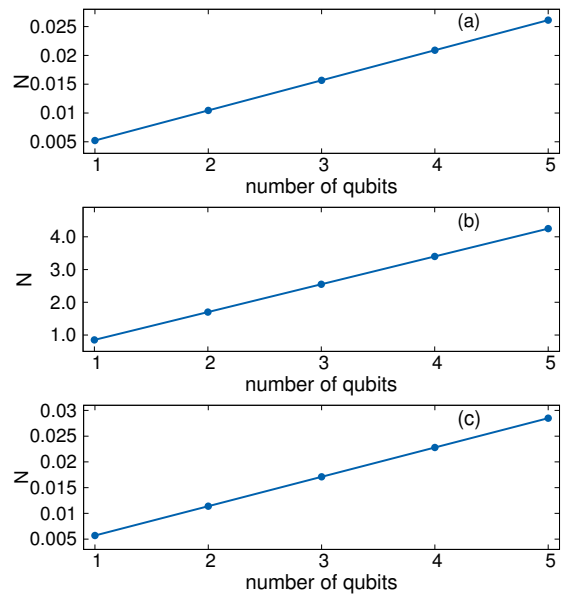} {}
\caption{Non-Markovianity ${\mathcal N(\Lambda)}$ for (a) the superohmic dephasing process with $s=3$, $w_c=1$ and $\eta=2$, for (b) the relaxation process with $\gamma_0=1$ and $\lambda=0.1$, and for (c) impurity atoms coupled to a BEC environment with $\sigma=45$ nm, as a function of the number of qubits.}
\end{figure}

Performing the optimization required for the calculation of the LFS measure ${\mathcal N(\Lambda)}$ becomes a very difficult task for a system of two or more qubits due to the significantly increasing number of variables involved in such cases. To overcome this difficulty, we limit our study to diagonal product initial states of the form $\rho^s(0)^{\otimes n}$ for multipartite systems. Indeed, this is a reasonable choice given the results acquired for the case of a single qubit. Moreover, by using such diagonal states we are able to obtain a lower bound for the degree of non-Markovianity of a quantum process. While we plot the LFS measure as a function of the number of qubits for superohmic dephasing environments with $s=3$, $w_c=1$ and $\eta=2$ in Fig. (3-a), similar results are displayed for the relaxation process with $\gamma_0=1$ and $\lambda=0.1$ in Fig. (3-b) and for the BEC environment with $\sigma=45$ nm in Fig. (3-c). These figures indicate a linear increase in the degree of non-Markovianity for the considered initial states, which proves that, for independent environments, the LFS measure is \textit{at least} additive, \textit{i.e.},
\begin{equation}
{\mathcal N}(\Lambda^{\otimes n}) \ge n[{\mathcal N}(\Lambda)].
\end{equation}
This peculiar behavior of the LFS measure, despite non-intuitive, is simple to be understood in this context because both the system and the environment entropies are additive quantities when independent environments are considered.


\subsection{Two-Qubit: Common Environment}

In this section, we turn our attention to the behavior of non-Markovianity when the system is globally interacting with a common environment. Due to the difficulty of calculating the dynamics for many qubits considering their interaction with a common environment, we restrict our analysis to only two qubits. However, the analysis of the two-qubit case is still interesting to infer the main characteristics of the degree of non-Markovianity as a function of the system scale. We first consider the case where both qubits are coupled to a common dephasing bath described by the following Hamiltonian
\begin{equation}
H_{CPD}=\frac{1}{2}\sum_{n} \omega_0^{(n)}\sigma_z^{(n)}+\sum_{k}\omega_k a_k^\dagger a_k+\sum_{n,k}\sigma_z^{(n)}(g_k a_k^\dagger + g_k^* a_k),
\end{equation}
where the index $n$ denotes the terms related to the first ($n=1$) and second ($n=2$) qubit. We also only consider diagonal two-qubit initial states to evaluate the degree of non-Markovianity measured by the LFS measure. The results of our analysis suggest that the LFS measure ${\mathcal N(\Lambda)}$ suffers a very significant decay as compared to the single qubit case, having a value of the order of $10^{-7}$. Note that such a finding is clearly different from what is observed for independent environments, where the degree of non-Markovianity increases at least linearly. 

Next, we focus on the scenario where a system of two qubits globally interact with a common relaxation environment:
\begin{equation}
H_{CAD}=\sum_{n}\omega_0^{(n)}\sigma_{+}^{(n)}\sigma_{-}^{(n)}+\sum_{k}\omega_k a_k^\dagger a_k + \sum_{n} (\sigma_{+}^{(n)}B + \sigma_{-}^{(n)}B^\dagger).
\end{equation}
In this case our investigation reveals that, by assuming diagonal initial states, the degree of non-Markovianity is significantly amplified when compared to the single qubit case, approximately turning out to be ${\mathcal N(\Lambda)}\approx 6.21$. Interestingly, among the two-qubit diagonal states that we have considered, the optimal one is always the maximally mixed state, independently of the bath parameters. Such a finding is rather surprising because the optimal initial state is strongly dependent on the parameters of the environment for the single qubit case.

Comparing the results obtained for the superohmic dephasing process with those obtained for the relaxation process, one could believe that while the degree of non-Markovianity increases for relaxation processes, it decreases for dephasing processes, as compared to the single qubit case. However, as shown below, this is not generally true. Although the situation where two impurity atoms are coupled to a BEC environment does not involve any energy exchange between the system and its surroundings, the degree of non-Markovianity still increases when compared to the case of a single qubit. In order to demonstrate this behavior, we first note that ${\mathcal N(\Lambda)}\approx 0.0055$ for a single qubit. Next, we calculate the degree of non-Markovianity for two qubits by taking the distance between the pair of impurity atoms in Eq. (\ref{gam2}) as $D=600$ nm. 
In this case, BEC environment interacts collectively with the pair of impurity atoms and we find ${\mathcal N(\Lambda)}\approx 0.0260$. This result corroborates with those obtained in Ref. \cite{sabrina2}, where the degree of non-Markovianity is studied considering the BLP measure. In fact, such an outcome suggests that the amount of non-Markovianity is not only connected to the exchange of energy between the system and environment but also intimately related to the spectral density of the reservoir modes and the dynamics of the coherence terms.


\section{Degree of Non-Markovianity: BLP Measure}
In this section, we aim to study the degree of non-Markovianity quantified by the BLP measure. Since this measure has been broadly studied in the literature, we repeat some of the earlier results \cite{BLP,haikkabec,sabrina2} here for the purpose of completeness of our work. Given the difficulty of calculating the BLP measure even numerically, we avoid the analysis of multi-qubit systems and focus on the cases of having one- and two-qubits systems separately.

\subsection{Single Qubit}
We begin our investigation by examining a single qubit system interacting with a reservoir. This problem has been first addressed in Ref. \cite{dephprob} for the dephasing process, in Ref. \cite{BLP} for the relaxation process, and in Ref. \cite{haikkabec, sabrina2} for the case of impurity atoms coupled to a BEC environment. We maintain the same parameters that we have used in the previous sections, \textit{i.e.}, $s=3$, $\omega_c=1$, and $\eta=2$ for the superohmic dephasing channel, $\gamma_0=1$ and $\lambda=0.1$ for the amplitude damping channel, and $\sigma=45$ nm for impurity atoms coupled to a BEC environment. The results for the BLP measure for a single-qubit considering different types of quantum processes are displayed in Table 1. These results will be useful
in order to understand the BLP measure for a pair of qubits, as discussed bellow.

\vspace{0.5cm}
\begin{minipage}{\linewidth}
%
\begin{tabular}{ C{2.25in} C{.5in} *2{C{.25in}}}\toprule[1.25pt]
\bf QUANTUM PROCESS & $\bf {\mathcal N}_{BLP}$ & \\\midrule
Dephasing        &  0.0432     & \\
Amp. Damping        &  0.9463     & \\
BEC Environment        &  0.0019     & \\
\bottomrule[1.25pt]
\end {tabular}\par
\medskip
TABLE 1: Degree of non-Markovianity of a single qubit \\ for different \;kinds of\; quantum\; processes\; considering \\the BLP measure.
\end{minipage}
\vspace{0.3cm}

\subsection{Two-Qubit: Independent Environments}
In this section, we employ the BLP measure to study the degree of non-Markovianity for two qubits that are independently interacting with uncorrelated environments. As already discussed in section IV.B, the amount of non-Markovianity quantified by the LFS measure increases at least linearly due to the additivity of the von-Neumann entropy. 
Following the same reasoning, our goal is to comprehend the degree of non-Markovianity quantified by the BLP measure for a system of two qubits. In Table 2, we present the results of our analysis for the superohmic dephasing process, the relaxation process, and for impurity atoms coupled to a BEC environment.

By comparing Table 1 and Table 2, we observe that the behavior of the degree of non-Markovianity, for independent environments, depends on the quantum process.
For the case of BEC environments, the BLP measure is \textit{approximately} $0.0019$ for one qubit, and $0.0038$ for two qubits subjected to independent environments.  As already shown in Ref. \cite{sabrina2}, the degree of non-Markovianity for this process is sub-additive. Indeed, the BLP measure also turns out to be sub-additive for independent relaxation environments. Furthermore, we emphasize that the outcomes of our analysis for the BLP measure do not generally agree with the results obtained for the LFS measure concerning the degree of non-Markovianity {for a single and a pair of qubits.} 
This feature is highlighted when we analyze the superohmic dephasing process. If we compare the single qubit case to the two qubits case subjected to independent superohmic dephasing environments, our numerical analysis points out an interesting result: the BLP measure remains constant for both cases. This result was verified by means of an exhaustive numerical analysis where we considered $10^6$ pairs of initial conditions encompassing both pure and mixed states.

\vspace{0.5cm}
\begin{minipage}{\linewidth}
%
\begin{tabular}{ C{2.25in} C{.5in} *2{C{.25in}}}\toprule[1.25pt]
\bf QUANTUM PROCESS & $\bf {\mathcal N}_{BLP}$ & \\\midrule
Dephasing        &  0.0432     & \\
Amp. Damping        &  1.2489     & \\
BEC Environment        &  0.0038    & \\
\bottomrule[1.25pt]
\end {tabular}\par
\medskip
TABLE 2: Degree of non-Markovianity of two qubits sub-\\jected to\; independent\; environments\; and different kinds of\\ quantum processes considering the BLP measure.
\end{minipage}
\vspace{0.3cm}

To understand the difference between both processes, we need to make a detailed analysis about the dynamics for one- and two-qubits subjected to independent environments.
For one qubit, the pair of states that maximize Eq. (\ref{blpmeas}) is given by $\rho_1(0)=|+\rangle$  and $\rho_2(0)=|-\rangle$ \cite{haikkabec}, where $|+\rangle$ and $|-\rangle$ are the eigenvectors of $\sigma_x$ Pauli matrix. In Fig. 4 we plot the density matrix coherence as a function of time, when the initial state is given by $\rho_1(0)$ and the quantum state is subjected to superohmic dephasing process (a) and to BEC environment (b). For both cases, we see that $\mathcal{N}_{BLP}$ is exactly given by $2\Delta$, where $\Delta$ is the amount of recoherence.

For two qubits subjected to superohmic independent environments, our numerical analysis show that the pair of states that maximize Eq. (\ref{blpmeas}) is given by $\rho_1(0)=|\!\downarrow\!+\rangle$ and $\rho_2(0)=|\!\downarrow\!-\rangle$, where $|\!\downarrow\rangle$ is one of the eigenvectors of the $\sigma_z$ Pauli matrix. In such a case, the only  non-zero coherence elements of the density matrix are $\rho_{34}$ and $\rho_{43}$ of the reduced density matrix,
\begin{eqnarray}
\rho=\left(\begin{array}{cccc}
\rho_{11} & \rho_{12} & \rho_{13} & \rho_{14}\\
\rho_{21} & \rho_{22} & \rho_{23} & \rho_{24}\\
\rho_{31} & \rho_{32} & \rho_{33} & \rho_{34}\\
\rho_{41} & \rho_{42} & \rho_{43} & \rho_{44}
\end{array}\right).
\end{eqnarray}
Furthermore, the dynamics of the element $\rho_{34}$ is the same as the one imposed to the coherence of one qubit. It means that the recoherence, and consequently the trace distance, are equivalent to the one qubit case, when the pair of states used to calculate the BLP are given by $\rho_1(0)=|\!\downarrow\!+\rangle$ and $\rho_2(0)=|\!\downarrow\!-\rangle$. Although the explanation for the superohmic dephasing process is clear, why the degree of non-Markovianity for the BEC environment increases? The fact that explains such behavior is related to a different pair of initial condition whose amount of recoherence is greater, $\rho_1(0)=\frac{1}{\sqrt{2}}(|\!\downarrow\uparrow\rangle-|\!\uparrow\downarrow\rangle)$ and $\rho_2(0)=\frac{1}{\sqrt{2}}(|\!\downarrow\uparrow\rangle+|\!\uparrow\downarrow\rangle)$. Thus, the amount of recoherence given by the element $\rho_{23}$ is greater than the amount of recoherence given by the element $\rho_{34}$ and consequently the BLP measure for the BEC model is bigger than the BLP measure for the superohmic dephasing model.

\begin{figure}
\includegraphics[width=.42\textwidth]{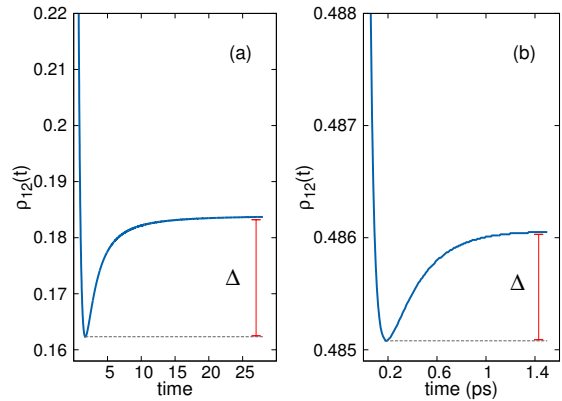} {}
\caption{Dynamics of the coherence term for (a) one qubit as a function of time for the superohmic dephasing process with $s=3$, $w_c=1$ and $\eta=2$, and for (b) the impurity atoms coupled to a BEC environment with $\sigma=45$ nm.}
\end{figure}

\subsection{Two-Qubit: Common Environment}
We explore the situation where two qubits are collectively interacting with a common environment. We once again consider the superohmic dephasing and the relaxation processes supposing a totally correlated environment, and the case of two qubits coupled to a BEC assuming that the distance between the pair of impurity atoms to be equal to $D=600$ nm. In such a configuration, the BEC interacts with the two qubits with a higher degree of correlation and thus we practically simulate a common environment. In Table 3, we display the results obtained for all three quantum processes.

\vspace{0.5cm}
\begin{minipage}{\linewidth}
%
\begin{tabular}{ C{2.25in} C{.5in} *2{C{.25in}}}\toprule[1.25pt]
\bf QUANTUM PROCESS & $\bf {\mathcal N}_{BLP}$ & \\\midrule
Dephasing        &  0.0002     & \\
Amp. Damping        &  7.8320     & \\
BEC Environment        &  0.0106    & \\
\bottomrule[1.25pt]
\end {tabular}\par
\medskip
TABLE 3: Degree of non-Markovianity for two qubits sub-\\jected to common environments and different kinds of quan-\\ tum processes considering the BLP measure.
\end{minipage}
\vspace{0.3cm}

Taking into account the values displayed in Table 1 and Table 3, we see that non-Markovianity quantified by the BLP measure decreases for the superohmic dephasing process when compared to the single qubit case. On the other hand, the BLP measure increases for both relaxation process and the case of impurity atoms coupled to a BEC environment. Particularly, the BLP measure is super-additive for these two processes.
In fact, such findings confirm that the amount of non-Markovianity is not only related to the exchange of energy between the system and the environment but also fundamentally connected to the spectral density of the reservoir modes and the dynamics of the coherence terms.


\section{Summary}

We study the degree of non-Markovianity of independent and common dephasing and relaxation processes for {a single and a pair of qubits}. 
Whereas we utilize the amplitude damping channel to represent dissipative processes, we consider the superohmic dephasing channel and the case of impurity atoms interacting with a BEC environment to describe phase damping processes. We develop our investigation by analyzing two conceptually different measures of non-Markovianity, a recently introduced quantity called the LFS measure, and the well known BLP measure.

Considering zero temperature environments, we show that no ancillary system is required to evaluate the degree of non-Markovianity quantified by the LFS measure since the quantity ${\mathcal N(\Lambda)}$ can be directly calculated by the difference of the time derivatives of the system and the environment entropies. Such a simplification provides an efficient method for calculating the degree of non-Markovianity due to the fact that the Hilbert space, where the maximization is evaluated, does not include an additional ancillary system. We provide an extensive analysis of the LFS measure for a single qubit and determine the optimal initial states of the system required for the evaluation of this particular measure, as a function of the parameters of the environment.

When it comes to the degree of non-Markovianity for independent environmental interactions, we demonstrate that the LFS measure might indeed \textit{increase} with the number of the qubits in the system for all considered quantum processes. In particular, we obtain a lower bound to the LFS measure for multipartite systems, namely ${\mathcal N}(\Lambda^{\otimes n}) \ge n[{\mathcal N}(\Lambda)]$ which implies that the LFS measure is at least additive. On the other hand, we have found the BLP measure to be sub-additive for the relaxation process and impurity atoms coupled to a BEC environment. More interestingly, for the superohmic dephasing process, our numerical analysis suggests that the BLP measure remains invariant, independent of whether we consider a system consisting of one qubit or two qubits.

Furthermore, we examine the behavior of non-Markovianity for a system of two qubits interacting with a common reservoir. In this scenario, the LFS and BLP measures agree on the general behavior of non-Markovianity. Particularly, the degree of non-Markovianity for the relaxation process is found to be super-additive for both of the measures. However, for a common environmental interaction, depending on the considered quantum process, the amount of non-Markovianity can be amplified or diminished as compared to the case of a single qubit. In fact, although the superohmic dephasing process and the case of impurity atoms coupled to a BEC environment do not involve any exchange of energy between the system and its surroundings, both LFS and BLP measures indicate that the degree of non-Markovianity is fundamentally different for these two processes. That is, while the degree of non-Markovianity for two impurity atoms coupled to a BEC environment is super-additive, the amount of non-Markovianity for the superohmic process decreases very significantly when compared to a single qubit. Indeed, such an outcome points out to the fact that the degree of non-Markovianity is significantly dependent on the spectral density of the reservoir modes and dynamics of the coherence terms.



\begin{acknowledgments}
We thank Mauro Paternostro, Bogna Bylicka, L. G. E. Arruda and Sabrina Maniscalco for fruitful discussions. This work is supported by FAPESP and CNPq through the National Institute for Science and Technology of Quantum Information (INCT-IQ) and by the Scientific and Technological Research Council of Turkey (TUBITAK) under Grant 111T232.
\end{acknowledgments}


\appendix
\section{Comparison of ${\mathcal N}(\Lambda)$ and ${\mathcal N_0}(\Lambda)$ }

\begin{figure} [htbp]
\includegraphics[width=.45\textwidth]{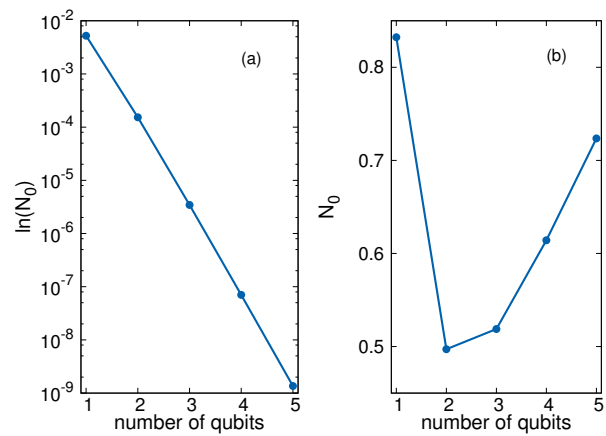} {}
\caption{Logarithm of non-Markovianity $\ln({\mathcal N_0(\Lambda)})$ for (a) the dephasing process with $s=3$, $w_c=1$ and $\eta=2$, and non-Markovianity ${\mathcal N_0(\Lambda)}$ for (b) the relaxation process with $\gamma_0=1$ and $\lambda=0.1$, as a function of number of qubits.}
\end{figure}

We compare the results obtained for the LFS measure ${\mathcal{N}}(\Lambda)$ and its simplified version ${\mathcal{N}_0}(\Lambda)$ for multipartite systems, considering independent environments. While the former involves a difficult maximization over all possible initial states, the latter can be directly calculated choosing a specific initial state, which we choose as a GHZ type state. Due to this restriction, it is clear that ${\mathcal{N}_0}(\Lambda)$ underestimates the degree of non-Markovianity, and consequently ${\mathcal N}_0(\Lambda)\le\mathcal{N}(\Lambda)$. In Fig. (5-a) we plot the logarithm of the simplified LFS measure, $\ln({\mathcal N_0(\Lambda)})$, as a function of the number of qubits for the superohmic dephasing process with $s=3$, $w_c=1$ and $\eta=2$. It can be observed that the degree of non-Markovianity measured by $\mathcal{N}_0(\Lambda)$ decays \textit{exponentially} as a function of the number of qubits. Furthermore, we see that even for very small systems (two qubits), the degree of non-Markovianity diminishes very significantly. In Fig. (5-b) we make the same analysis for the relaxation process considering the parameters $\gamma_0=1$ and $\lambda=0.1$. Our findings demonstrate that, unlike in the case of superohmic dephasing, the simplified LFS measure ${\mathcal N}_0(\Lambda)$ might increase for the relaxation process. As a result, comparing Fig. (5-a) and Fig. (5-b) to Fig. (3-a) and Fig. (3-b), we conclude that ${\mathcal N}_0(\Lambda)$ might be a misleading quantity for determining the degree of non-Markovianity, despite the fact that it is an easily computable witness of non-Markovianity.


\begin{thebibliography}{99}
\bibitem{general} M. M. Wolf, J. Eisert, T. S. Cubitt, and J. I. Cirac, Phys. Rev. Lett. \textbf{101}, 150402 (2008);
X.-M. Lu, X. Wang, and C. P. Sun, Phys. Rev. A \textbf{82}, 042103 (2010);
Z. Y. Xu, W. L. Yang, and M. Feng, Phys. Rev. A \textbf{81}, 044105 (2010);
R. Vasile, S. Maniscalco, M. G. A. Paris, H.-P. Breuer, and J. Piilo, Phys. Rev. A \textbf{84}, 052118 (2011);
B.-H. Liu, L. Li, Y.-F. Huang, C.-F. Li, G.-C. Guo, E.-M. Laine, H.-P. Breuer, and J. Piilo, Nat. Phys. \textbf{7}, 931 (2011);
P. Haikka, J. Goold, S. McEndoo, F. Plastina, and S. Maniscalco, Phys. Rev. A \textbf{85}, 060101(R) (2012);
A. Chiuri, C. Greganti, L. Mazzola, M. Paternostro, and P. Mataloni, Sci. Rep. \textbf{2}, 968 (2012);
L. Mazzola, C. A. Rodriguez-Rosario, K. Modi, and M. Paternostro, Phys. Rev. A \textbf{86}, 010102(R) (2012).
\bibitem{haikkabec} P. Haikka, S. McEndoo, G. De Chiara, G. M. Palma, and S. Maniscalco, Phys. Rev. A \textbf{84}, 031602(R) (2011).
\bibitem{sabrina2} C. Addis, P. Haikka, S. McEndoo, C. Macchiavello, and S. Maniscalco, Phys. Rev A. \textbf{87} 052109 (2013).
\bibitem{bogna} B. Bylicka, D. Chru\'{s}ci\'{n}ski, and S. Maniscalco, arXiv:1301.2585.
\bibitem{metro} A. W. Chin, S. F. Huelga, and M. B. Plenio, arXiv:1103.1219.
\bibitem{key} R. Vasile, S. Olivares, M. G. A. Paris, and S. Maniscalco, Phys. Rev. A \textbf{83}, 042321 (2011).
\bibitem{size} M. \v{Z}nidari\v{c}, C. Pineda, and I. Garc\'{\i}a-Mata, Phys. Rev. Lett. \textbf{107}, 080404 (2011);S. Lorenzo, F. Plastina, and M. Paternostro, arXiv:1205.4535.
\bibitem{onetwo} X. M. Lu, X. Wang and C. P. Sun, Phys. Rev. A. \textbf{82}, 042103 (2010); Z. Y. Xu, W. L. Yang, and M. Feng, Phys. Rev. A \textbf{81}, 044105 (2010); E.-M. Laine, H.-P. Breuer, J. Piilo, C.-F. Li, and G.-C. Guo, Phys. Rev. Lett. \textbf{108}, 210402 (2012).
\bibitem{BLP} H.-P. Breuer, E.-M. Laine, J. Piilo, Phys. Rev. Lett. \textbf{103}, 210401 (2009).
\bibitem{RHP} A. Rivas, S.F. Huelga, and M.B. Plenio, Phys. Rev. Lett. \textbf{105}, 050403 (2010).
\bibitem{Luo} S. Luo, S. Fu, and H. Song, Phys. Rev. A \textbf{86}, 044101 (2012).
\bibitem{other} E.-M. Laine, J. Piilo, and H.-P. Breuer, Phys. Rev. A \textbf{81}, 062115 (2010); L. Mazzola, E.-M. Laine, H.-P. Breuer, S. Maniscalco, and J. Piilo, Phys. Rev. A \textbf{81}, 062120 (2010); S. C. Hou, X. X. Yi, S. X. Yu, and C. H. Oh, Phys. Rev. A \textbf{83}, 062115 (2011).
\bibitem{soldeph} G. M. Palma, K.-A. Suominen, and A. K. Ekert, Proc. R. Soc. London A \textbf{452}, 567 (1996); J. H. Reina, L. Quiroga, and N. F. Johnson, Phys.Rev.A \textbf{65}, 032326 (2002).
\bibitem{Cirone} M. A. Cirone, G. De Chiara, G. M. Palma, and A. Recati, N.
Jour. Phys. \textbf{11}, 103055 (2009).
\bibitem{dephprob} Z. He, J. Zou, L. Li, and B. Shao, Phys. Rev. A \textbf{83}, 012108 (2011).
\end{thebibliography}
\end{document}